\documentstyle[aps,prl,preprint,floats,epsfig,subfigure]{revtex}

\begin{document}

\preprint{\tighten\vbox{\hbox{\hfil CLNS 01/1748}
                        \hbox{\hfil CLEO 01-15}
}}

\title{Evidence for the Decay $D^0\to K^+\pi^-\pi^+\pi^-$}  

\author{CLEO Collaboration}
\date{August 10, 2001}

\maketitle
\tighten

\begin{abstract} 
We present a search for the ``wrong-sign'' decay
$D^0 \to K^+\pi^-\pi^+\pi^-$ using $9~\mathrm{fb}^{-1}$ of $e^+e^-$
collisions on and just below the $\Upsilon(4S)$ resonance.  This decay
can occur either through a doubly Cabibbo-suppressed process or through
mixing to a $\overline{D}^0$ followed by a Cabibbo-favored process.  Our
result for the time-integrated wrong-sign rate relative to the decay
$D^0 \to K^-\pi^+\pi^-\pi^+$ is
\mbox{
$(0.0041^{+0.0012}_{-0.0011}(\mathrm{stat.})\pm0.0004(\mathrm{syst.}))\times(1.07\pm0.10)(\mathrm{phase~space})$,
}
which has a statistical significance of 3.9 standard deviations.
\end{abstract}
\newpage

{
\renewcommand{\thefootnote}{\fnsymbol{footnote}}

\begin{center}
S.A.~Dytman,$^{1}$ V.~Savinov,$^{1}$
S.~Chen,$^{2}$ J.~W.~Hinson,$^{2}$ J.~Lee,$^{2}$
D.~H.~Miller,$^{2}$ E.~I.~Shibata,$^{2}$ I.~P.~J.~Shipsey,$^{2}$
V.~Pavlunin,$^{2}$
D.~Cronin-Hennessy,$^{3}$
% ...
\end{center}
 
\small
\begin{center}
$^{1}${University of Pittsburgh, Pittsburgh, Pennsylvania 15260}\\
$^{2}${Purdue University, West Lafayette, Indiana 47907}\\
$^{3}${University of Rochester, Rochester, New York 14627}\\
% ...
\end{center}

\setcounter{footnote}{0}
}
\newpage

Mixing between $D^0$ and $\overline{D}^0$ mesons occurs because they are
eigenstates of the strong interaction Hamiltonian but not of the total
Hamiltonian.
The Standard Model predicts a very small rate of
mixing~\cite{theory:predictions}, though new physics processes can enhance
the effect~\cite{theory:new_phys}.
At present, the preferred mode in which to search for mixing is
$D^0 \to K^+\pi^-$~\cite{cleo:kpi,focus:kpi},\footnote{Charge conjugate
reactions are implied throughout this paper.} since it tends to have a high
signal-to-background
ratio, and its proper time distribution can be directly interpreted in terms
of the rates for mixing and for doubly Cabibbo-suppressed~\cite{theory:ckm}
decays (DCSD).
The best limits on the rate of mixing relative to the Cabibbo-favored
decay ($R_{mix} < 0.2\%$ at the 95\% confidence level~\cite{cleo:kpi}) are
obtained from this decay mode.
Other ``wrong-sign'' modes of the form $D^0 \to K^+(n\pi)^-$ are
interesting to study, as they are also sensitive to mixing and
present us with additional information about the doubly
Cabibbo-suppressed decays which interfere with mixing~\cite{cleo:kpipi0}.
Without lifetime information, the best limits on $R_{mix}$ come from
semileptonic decay modes, where any wrong-sign signal is evidence for
mixing because there is no DCSD contribution ($R_{mix} < 0.5\%$ at the
90\% confidence level~\cite{e791:semilep}), and the backgrounds are the
limiting factor.

In this paper we present the ratio of branching fractions
\begin{displaymath}
   R_{WS} \equiv
   \frac{{\mathcal B}(D^0 \to K^+\pi^-\pi^+\pi^-)}
        {{\mathcal B}(D^0 \to K^-\pi^+\pi^-\pi^+)}
   = R_D + \sqrt{R_D}\;y' + R_{mix},
\end{displaymath}
where $R_D$ is the relative rate for the doubly Cabibbo-suppressed decay with
respect to the Cabibbo-favored decay,
and $y'$ is the imaginary part
of the mixing amplitude after rotation through the strong phase $\delta$,
which depends on the specific final state.
In $D^0 \to K^+\pi^-$, $\delta_{K\pi}$ is just the phase difference between
the Cabibbo-favored and doubly Cabibbo-suppressed decay amplitudes.
In $D^0 \to K^+\pi^-\pi^+\pi^-$, there can be many resonant sub-modes which
interfere.
After integrating over the phase space, this results in a total phase
difference $\delta_{K3\pi}$ which is analogous to $\delta_{K\pi}$ but
is difficult to interpret.

The potentially different phase space structures of right-sign and wrong-sign
add an additional complication in measuring the ratio of branching fractions.
If the reconstruction efficiencies differ between the two, then $R_{WS}$
will be:
\begin{equation}
   R_{WS} =
   \frac{N(D^0 \to K^+\pi^-\pi^+\pi^-)}{N(D^0 \to K^-\pi^+\pi^-\pi^+)}
   \times
   \frac{\left\langle\varepsilon_{RS}\right\rangle}
        {\left\langle\varepsilon_{WS}\right\rangle}
   \label{eq:rws}
\end{equation}
where $N$ is the number of events observed, and
$\left\langle\varepsilon\right\rangle$ is the reconstruction efficiency
for right-sign or wrong-sign, averaged over phase space.  We also define
\begin{displaymath}
   r \equiv \frac{N(D^0\to K^+\pi^-\pi^+\pi^-)}{N(D^0\to K^-\pi^+\pi^-\pi^+)}.
\end{displaymath}

The data used in this analysis consist of an integrated luminosity of
$9~\mathrm{fb}^{-1}$ of $e^+e^-$ collisions collected with the CLEO II.V
detector~\cite{cleoii:nim,cleoiiv:si,cleoiiv:dr} between 1996 and 1999.
The detector is located in Ithaca, NY at the Cornell Electron Storage Ring,
which collides electrons and positrons at a center of mass energy on
or just below the $\Upsilon(4S)$ resonance.
These collisions take place in a luminous region at the center of the
detector with a Gaussian width of approximately $10~\mu$m in the vertical
direction, $300~\mu$m in the horizontal direction, and 1~cm along the beam
direction~\cite{cesr:luminous}.

The detector consists of a number of concentric cylindrical sub-detector
components immersed in a 1.5~T solenoidal magnetic field oriented along the
axis, which is the direction of the incident beams.
The innermost component is a silicon vertex detector, a high-precision
tracking device with three layers of silicon wafers, instrumented on
both sides, which measures particle position to 30~$\mu$m in the $r-\phi$
plane and 100~$\mu$m along the beam direction.
Immediately outside of this are two drift chambers with a total of 50 layers
of axial wires and 11 layers of stereo wires.
The position resolution for each layer in the drift chambers is approximately
100~$\mu$m.
There are additional components at larger radii, including time-of-flight
counters, an electromagnetic calorimeter, and muon detectors, but these
are not used in this analysis.

Monte Carlo simulated events are used to determine selection criteria,
obtain fitting shapes, and study some of the systematic effects.
The simulation is based on the GEANT package~\cite{geant}, and its
output is processed in the same manner as the data.
The size of the Monte Carlo sample is equivalent to approximately eleven
times the size of the data sample.

A $D^0$ candidate is fully reconstructed from a charged kaon and three charged
pions ($K3\pi$).
The initial flavor of the $D^0$ candidate is determined by the charge of
the slow pion, $\pi_s^+$, from the decay chain $D^{*+} \to D^0 \pi_s^+$.
A right-sign event is one in which the charge of the kaon
is opposite from the charge of the slow pion, while a
wrong-sign event is one in which they are the same~\cite{marsh:thesis}.

The $D^0$ candidate is subject to a number of requirements.
The decay particles are constrained to come from a common vertex, which
eliminates candidates constructed of some particles originating from
a displaced vertex and others from either another displaced vertex or
the primary interaction point.
The resulting invariant mass distribution has a width of 5.4~MeV/$c^2$.
There are many more right-sign than wrong-sign events, so we must
remove fake wrong-sign events due to mis-identified right-sign events,
or feed-through, which can bias our signal.
We want to treat both the right-sign and wrong-sign samples identically,
so we must also remove feed-through in the other direction.
Feed-through occurs when the kaon is mis-identified as a pion, and one
of the oppositely charged pions is mis-identified as a kaon.
To determine if a candidate is feed-through, we swap two of our mass
assignments such that the decay switches between ``right'' and ``wrong''
sign.
From this we calculate an invariant mass and its uncertainty, and
if the invariant mass is less than 3.85 standard deviations from the
nominal $D^0$ mass, the candidate is considered feed-through.
In addition, we require consistency
between the observed and expected rate of energy loss ($dE/dx$) in the
wire chambers given the observed momentum of a particle and its mass
hypothesis.
Kaon candidates are required to have energy loss within 2.1 standard
deviations of the expected value; pion candidates within 3.2 standard
deviations.

We constrain the slow pion candidates to pass through the intersection
of the $D^0$ trajectory and the vertical position of the center
of the luminous region.
This constraint removes much of the mis-measurement of the slow pion due to
multiple Coulomb scattering in the beam pipe and the first layer of the
silicon detector.
The improved slow pion measurement results in a 210~keV-wide distribution
of the energy release of the $D^{*+}$ candidate,
\begin{displaymath}
   Q \equiv m(K3\pi+\pi_s) - m(K3\pi) - m_\pi,
\end{displaymath}
where $m(K3\pi)$ is the invariant mass of the $D^0$ candidate, and
$m(K3\pi+\pi_s)$ is the invariant mass of the $D^{*+}$ candidate.
A minimum momentum of 2.68~GeV/$c$ is required of the
$D^{*+}$ candidate to reduce the combinatoric background; it
also kinematically removes decays originating from $B$ mesons.

One particular background, which we must explicitly veto, is the singly
Cabibbo-suppressed
decay $D^0 \to K^+ \bar{K}^0 \pi^-$ followed by $\bar{K}^0 \to \pi^+\pi^-$.
The combined branching fraction for this process is
$(1.7\pm0.3)\times10^{-3}$~\cite{e691:k0bar,cleo:k0bar}, and it can
masquerade as our wrong-sign signal.
Much of this will be excluded by the common-vertex constraint on the $K3\pi$,
however, what remains will look identical to the signal.
The reconstructed invariant mass of the $\pi^+\pi^-$ from the $\bar{K}^0$
will be shifted down slightly from the $K_S^0$ mass because the pion
tracks will tend to be pulled back towards the $D^0$ decay point
by the common-vertex constraint.
In order to eliminate these decays, we require that neither $\pi^+\pi^-$
combination have an invariant mass between 472~MeV/$c^2$ and
512~MeV/$c^2$.  This requirement is 90\% efficient for the right-sign
signal.

\begin{figure}[ht]
   \begin{center}
      \mbox{
         \subfigure[$K3\pi$ invariant mass]{
            \epsfig{file=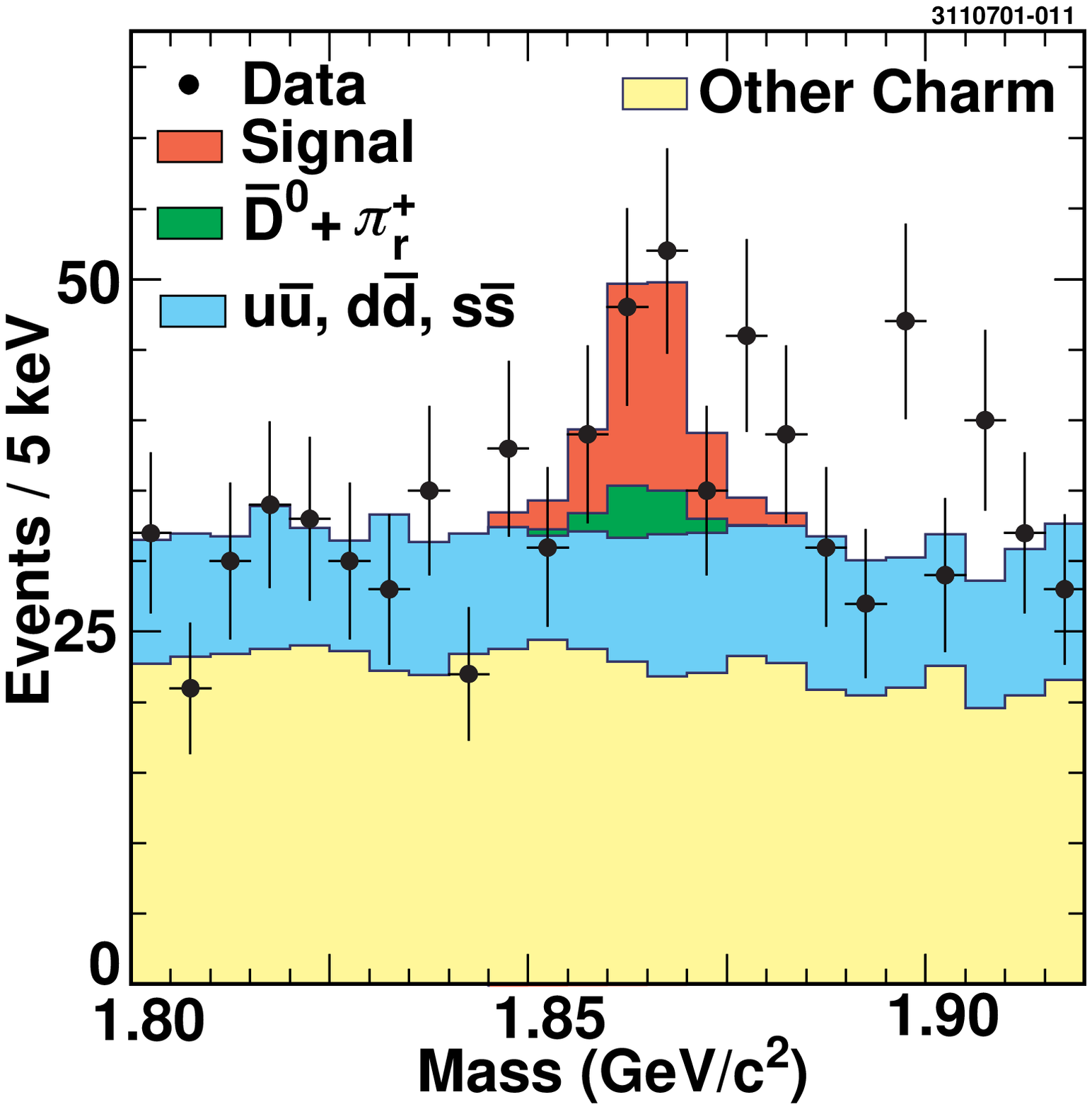,width=0.45\textwidth}
         }
         \quad
         \subfigure[Energy release of $K3\pi+\pi_s$]{
            \epsfig{file=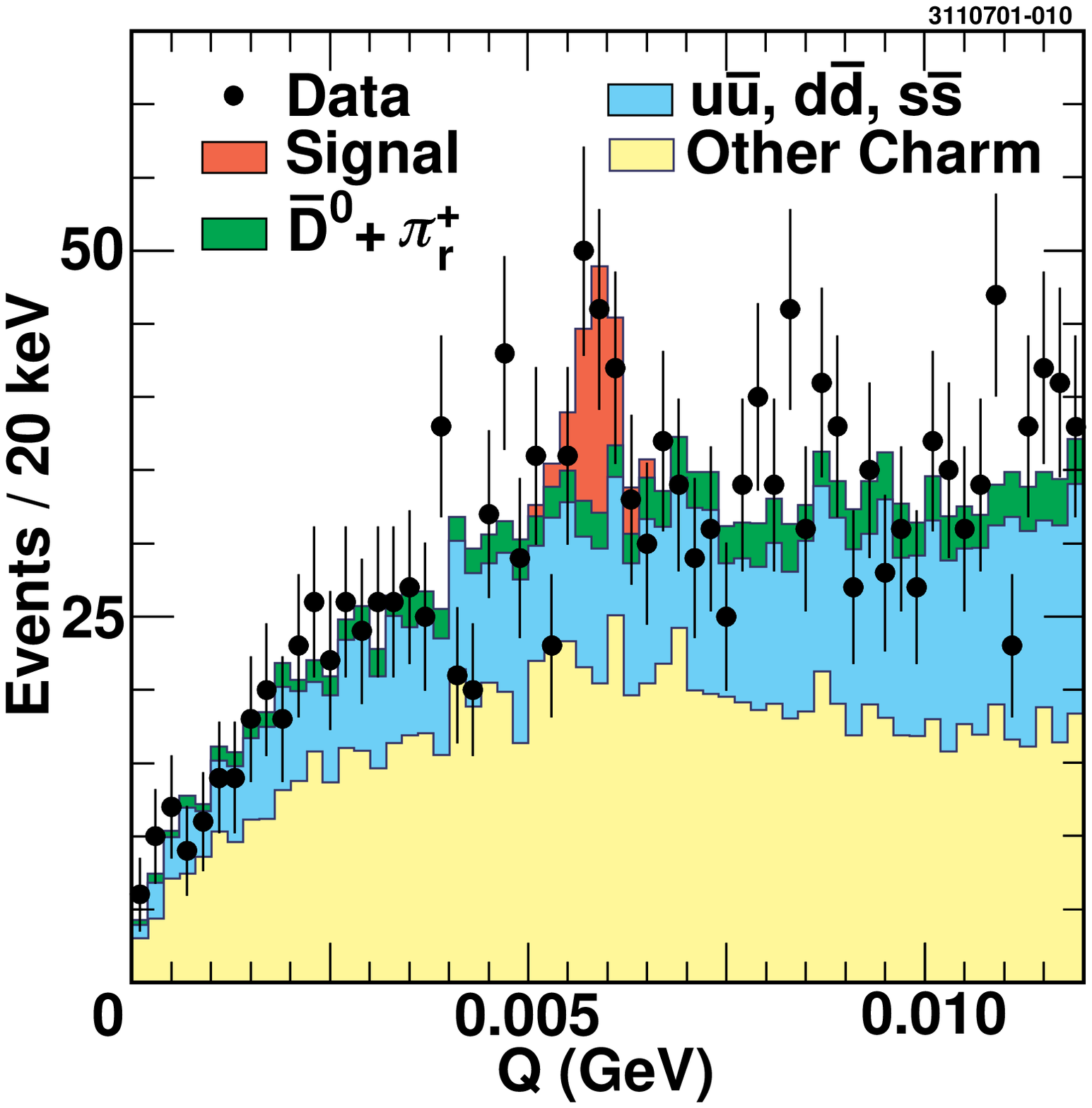,width=0.45\textwidth}
         }
      }
      \caption{Projections of the 2D distributions on the (a) $m(K3\pi)$
               and (b) $Q$ axes, after selecting events within two standard
               deviations of the signal peak in the other variable.  The data,
               signal, and three background components are shown.}
      \label{fig:projections}
   \end{center}
\end{figure}

We determine the ratio $r$ by performing a two-dimensional binned likelihood
fit of the wrong-sign data in the plane of $Q$ {\em vs.}\ $m(K3\pi)$.
Histograms are used as the fitting shapes for the signal and backgrounds
rather than analytic forms in order to simplify the fits.
The signal shape histogram comes from the right-sign data, reducing any
possibility of mis-modeling the shape of the signal in the wrong-sign data.
This has the additional benefit that we can fit directly for $r$.
A simple fit of the right-sign $m(K3\pi)$ distribution to a constant plus
two Gaussians yields a signal-to-background ratio of about 35:1 in the
signal region, which is low enough that we neglect the effect of the
background in the signal shape.
The presence of background in the signal shape does not bias $r$, since
the right-sign background is scaled along with the signal and does not
appreciably alter the shape of the distribution.
The backgrounds consist of three categories:
true right-sign $\overline{D}^0 \to K3\pi$ decays paired with a
random $\pi_s^+$,
other $e^+e^- \to c\bar{c}$ events,
and events originating from $e^+e^- \to u\bar{u},d\bar{d},s\bar{s}$.
The first background produces a peak in the $K3\pi$ invariant mass, but not
in the $Q$ distribution.
The second is flat in mass, but produces a very broad peak in $Q$ due to
partially reconstructed $D^0$ decays matched with a slow pion from a
real $D^{*+}$.
The third does not produce a peak in either mass or $Q$.
For all three categories of background, we obtain fitting histograms
from the Monte Carlo simulation.
The normalizations of the signal shape and all three background shapes are
allowed to vary independently in the fit.
The fit result for $r$ is $0.0041^{+0.0012}_{-0.0011}(\mathrm{stat.})$,
which has a statistical significance of 3.9 standard deviations.
The right-sign data contain 13735 events, which yields $54\pm14$
wrong-sign signal events using $r$ and accounting for the small
background in the right-sign signal shape.
Projections of the two-dimensional fit onto the mass and $Q$ axes for
combinations  within
two standard deviations of the central value of the other variable are
shown in Figure~\ref{fig:projections}.

With four particles in the final state, the phase space for the decay
is five-dimensional.
The reconstruction efficiency is not completely uniform across the phase space,
so if the right-sign and wrong-sign events populate phase space differently,
there will be a multiplicative correction to account for the
difference in average efficiencies, as indicated in Equation~(\ref{eq:rws}).
This correction is studied using the data and an additional sample of
simulated signal with a flat matrix element.
A small bubble in phase space is formed around each data event in the signal
region, in which we count the number of generated and reconstructed simulated
events.
This gives us the local efficiency for the data event, and its reciprocal is
the event weight.
The sum of these weights is the total number of expected right-sign
or wrong-sign events, and using the observed yields we calculate the
average efficiency.
The wrong-sign data have significant background, however, so we perform
a background subtraction to obtain the signal weight of the wrong-sign
data.
To calculate the background weight, we first find the total weight of the
wrong-sign events outside of the signal region in $Q$ but within the signal
region in mass.
This sideband weight is then scaled by the relative numbers of background
events in the signal region and the sidebands, and the difference between
the total wrong-sign weight and this background weight is the wrong-sign
signal weight.
The ratio of average efficiencies is found to be $1.07\pm0.10$. Our
precision on the efficiency ratio is limited by the size of the wrong-sign
signal sample in the data.

Many sources of systematic uncertainty cancel in measuring the ratio $r$.
The background under the wrong-sign peak is the dominant effect in determining
this ratio, so the quality of the simulation determines the most
significant sources of systematic uncertainty.
The relevant sources are
the statistical precision of the simulated sample
and
how well the simulation models the background shapes.
Of these, the statistical precision of the sample is the larger.

The shapes used in fitting the backgrounds are derived from a finite-size
simulated sample and so are subject to statistical fluctuations.
To account for this, we form an ensemble of shapes fluctuated bin-by-bin
according to a Poisson distribution.
The mean of the Poisson distribution for these fluctuations is the
contents of the bin from the full simulated sample, and each shape
is fluctuated separately.
The distribution for the values of $r$ obtained by fitting the data
with these fluctuated shapes is a Gaussian whose width corresponds to
an absolute systematic uncertainty on $r$ of 0.00032.
Using means that are already the results of statistical fluctuation
will tend to accentuate the statistical effects, and the uncertainty
from this systematic effect is estimated conservatively.
This is the largest systematic uncertainty, but is still considerably
smaller than the statistical uncertainty.

In addition to statistical variations, it is possible that the simulation
is mis-modeling the backgrounds.
In order to obtain a quantitative estimate of this effect, we perform the
fit with various regions outside of the signal region excluded.
This probes more local variations between the data and simulation than
the fit to the full $Q-m(K3\pi)$ plane.
Specifically, we perform eight separate fits, four to the quadrants of
the $Q-m(K3\pi)$ plane, and four to the half-planes formed by adjacent
quadrants.
In each case, we include the entire signal region.
The sample standard deviation of these restricted fits gives us a
systematic uncertainty of 0.00031.
Since the restricted fits are correlated with
the statistical precision of the full fit and each other, the systematic
uncertainty we assign is a conservative estimate.

Combining the systematic uncertainties discussed above in quadrature
gives $\pm0.0004$ for the total systematic uncertainty on $r$.
Our final result is
\[
R_{WS}=(0.0041^{+0.0012}_{-0.0011}(\mathrm{stat.})\pm0.0004(\mathrm{syst.}))
\times(1.07\pm0.10)(\mathrm{phase~space}).
\]
This result is a substantial improvement over the previous result of\linebreak
\mbox{
$R_{WS}\!=\!0.0025^{+0.0036}_{-0.0034}(\mathrm{stat.})\pm0.0003(\mathrm{syst.})$
}
from E791~\cite{e791:knpi}, for which the right-sign to wrong-sign phase-space
efficiency correction was assumed to be exactly 1.0.
Our result, with a statistical significance of 3.9 standard deviations,
provides guidance towards the luminosities that future experiments
will need in order to measure the mixing and DCSD phase-space structure
of the decay $D^0\to K3\pi$.

We gratefully acknowledge the effort of the CESR staff in providing us with
excellent luminosity and running conditions.
M. Selen thanks the PFF program of the NSF and the Research Corporation,
and A.H. Mahmood thanks the Texas Advanced Research Program.
This work was supported by the National Science Foundation, the
U.S. Department of Energy, and the Natural Sciences and Engineering Research
Council of Canada.

\end{document}